\documentclass[osajnl,twocolumn,showpacs,superscriptaddress,10pt]{revtex4-1} 
\usepackage{amsmath,amssymb,graphicx,epstopdf}
\begin{document}

\title{Observation of nitrogen vacancy photoluminescence from an optically levitated nanodiamond}


\author{Levi P. Neukirch}
\affiliation{Department of Physics and Astronomy, University of Rochester, Rochester, NY 14627, USA}

\author{Jan Gieseler}
\affiliation{ICFO-Institut de Ciencies Fotoniques, Mediterranean Technology Park, 08860 Castelldefels (Barcelona), Spain}

\author{Romain Quidant}
\affiliation{ICFO-Institut de Ciencies Fotoniques, Mediterranean Technology Park, 08860 Castelldefels (Barcelona), Spain}
\affiliation{ICREA-Instituci\'{o} Catalana de Recerca i Estudis Avan\c{c}ats, 08010 Barcelona, Spain}

\author{Lukas Novotny}
\affiliation{Photonics Laboratory, ETH Z\"{u}rich, 8093 Z\"{u}rich, Switzerland.}
\affiliation{Institute of Optics, University of Rochester, Rochester, NY 14627, USA}

\author{A. Nick Vamivakas}
\email{nick.vamivakas@rochester.edu}
\affiliation{Institute of Optics, University of Rochester, Rochester, NY 14627, USA}

\begin{abstract}
We present the first evidence of nitrogen vacancy (NV) photoluminescence from a nanodiamond suspended in a free-space optical dipole trap at atmospheric pressure. The photoluminescence rates are shown to decrease with increasing trap laser power, but are inconsistent with a thermal quenching process. For a continuous-wave trap, the neutral charge state (NV$^0$) appears to be suppressed. Chopping the trap laser yields higher total count rates and results in a mixture of both NV$^0$ and the negative charge state (NV$^-$).
\end{abstract}


\maketitle 

\noindent The negatively charged nitrogen vacancy center (NV$^-$) in diamond has drawn substantial interest in quantum optics, quantum information \cite{Wrachtrup2006processing,Neumannetal2010NaturePhys}, and nanoscale sensing \cite{Mazeetal2008NanoscaleMagneticSensing,Doldeetal2011Electric,Mamin2013NMR}. It has proven to be a stable source of single photons \cite{Kurtsieferetal2000g2}, and displays a long ground-state spin coherence lifetime at room temperature \cite{Stanwix2010PRB}. It has served as a stable, optically accessible qubit in bulk diamond \cite{Jelezkoetal2004}, and has been used to mediate spin reading and writing to $^{13}$C nuclei \cite{Maureretal2012stablespinlifetimes}. Recently nanodiamonds containing ensembles of NV centers \cite{Horowitzetal2012trappednanodiamonds}, and single defects \cite{Geiselmannetal2013trappednanodiamonds} have been demonstrated to be well suited to biological magnetic sensing applications in optical tweezers, and sample characterization using microscopy techniques such as fluorescence lifetime imaging microscopy (FLIM) \cite{Schelletal2013FLIM,Beamsetal2013FLIM}.

In this letter we report the first measurement of photoluminescence (PL) from a nanodiamond containing an ensemble of NV centers that is levitated in a free-space optical dipole trap. Such a system offers several advantages over previously demonstrated optical tweezer experiments. Most importantly, by eliminating the need for a liquid solution, our method represents a natural first step toward trapping nanodiamonds in vacuum, and implementing optomechanical cooling schemes \cite{Lietali2011Cooling, Gieseleretal2012} Optically levitated and cooled dielectric particles have been demonstrated to be superb optomechanical resonators, with extremely high mechanical quality factors \cite{Ashkin1976HighVacuum}. Furthermore, cooling the center-of-mass (COM) motion of these particles to their quantum ground state, while technically challenging appears feasible \cite{Gieseleretal2012}. Such a system using a nanodiamond containing an NV center, promises a hybrid quantum system of not only unprecedentedly high mechanical $Q$-factor, but one in which the $Q$-factor is tunable (via ambient pressure). A recent proposal suggests the possibility of generating Schr\"odinger cat states with an optically levitated nanodiamond \cite{Yinetal2013cats}.

\begin{figure}[h]
\begin{centering}
\includegraphics[width=7.5cm]{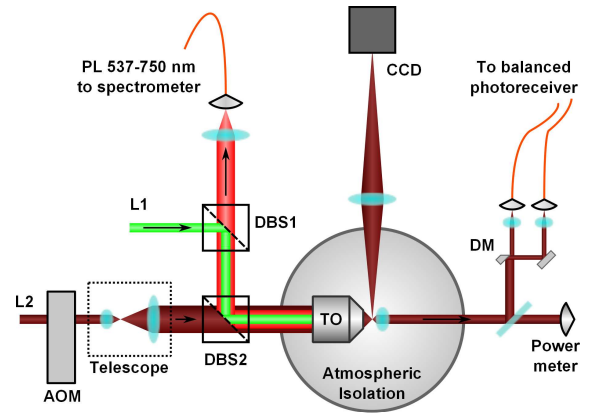}
\caption{The optical trap is formed in a vacuum chamber using the NA=0.9 trap objective (TO). The excitation laser (L1) and photoluminescence are combined with a longpass dichroic beamsplitter (DBS1, 560 nm edge), and then combined with the trap laser (L2) using a second dichroic beamsplitter (DBS2, 776 nm edge). The trap laser is chopped using an acousto-optic modulator (AOM) before being expanded by a lens pair.\label{fig:layout}}
\end{centering}
\end{figure}

The optical trap is formed by focusing a continuous wave Neodynium:YAG laser (1064 nm) using a NA=0.9 Nikon microscope objective [Fig. \ref{fig:layout}]. An acousto-optic modulator (AOM) is positioned in the trap beam, with the first-order refracted beam used for trapping. The AOM is driven by a delay/pulse generator with sub-ns precision (SRS DG535), allowing the trap laser to be chopped with rise/fall times of ~100 ns (limited by the AOM response). The trap beam is then expanded to fill the back aperture of the trapping objective. A Picoquant LDH-FA pulsed laser (532 nm; 40 MHz, 100 ps pulsewidth) is used to excite the NV centers. PL is collected into a multimode fiber. A 537 nm longpass filter positioned before the fiber prevents any excitation light from leaking into the fiber and exciting background fluorescence. A pair of longpass dichroic beamsplitters (edges at 560 nm and 776 nm) allows confocal alignment of all three optical channels. For these experiments, the fiber output passes though a 750 nm shortpass filter to remove any 1064 nm light before entering a 0.75 m long imaging spectrometer with a liquid nitrogen cooled CCD (Princeton Instruments).

Trapped particles are imaged from the side by focusing horizontally scattered laser light onto a CCD.  To measure the motion of a trapped particle, an aspheric lens (NA=0.68) is positioned after the trap to collimate the outgoing light. A beam sampler sends 10\% of the outgoing light to monitor the particle's transverse-horizontal position using a D-mirror and balanced photoreceiver \cite{GittesandSchmidt1998,Chavez2008RevSciInstr}. During position measurements the excitation laser is blocked before it reaches the chamber. These experiments were conducted at atmospheric pressure; however, to isolate the trap from air currents present in the laboratory, the trapping objective and collimating lens were housed in a cylindrical, 8" diameter, stainless steel chamber. The trapping optics were mounted in the chamber with off the shelf optomechanical components, and access for the laser beams and imaging was provided by optical-quality fused-silica windows.

\begin{figure}
\begin{centering}
\includegraphics[width=8.5cm]{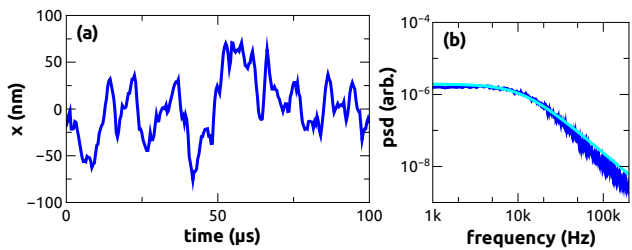}
\caption{At atmospheric pressure, (a) position measurements and (b) power spectra of position show the highly damped, Brownian nature of the motion of an optically levitated nanodiamond. A curve fit to the PSD according to Eq. (\ref{eq:PSD}) is shown in cyan, and results in an estimated particle radius of 38 nm.\label{fig:position}}
\end{centering}
\end{figure}

We use commercially available fluorescent HPHT nanodiamonds manufactured by Ad\'{a}mas Nanotechnologies (ND-NV-100nm). The crystals are specified to have an average size of 100 nm and contain $\sim500$ NV centers. The diamonds are used as shipped, with no subsequent processing to enhance their PL properties. To load the trap we mix a dilute solution ($\sim1000:1$) of the nanodiamond slurry with ethanol, and vaporize this solution with an ultrasonic nebulizer. The vapor is sprayed into the trapping chamber and we wait for diamonds to drift into the focal spot. This process typically results in a trapped particle after a few minutes.

Powers used for trapping ($P_{trap}$) usually range from 50-100 mW. Once trapped, most diamonds are stable in a CW or chopped trap for minutes to hours. An example of a typical position measurement is presented in Fig. \ref{fig:position}a. The output of the balanced photodetector is recorded via a computer (NI PCIe-6531 data acquisition card) or by a digital oscilloscope. At atmospheric pressure the harmonic nature of the trap is hidden by the high mechanical damping rate, and the motion is Brownian \cite{Lietal2010Brownian}. Power spectral densities of position show a high frequency roll-off consistent with overdamped motion (see Fig. \ref{fig:position}b). Still, it is possible to estimate the size of a trapped particle.

By making the dipole and paraxial approximations, the displacements of a particle of mass $m$ from trap center result in motion satisfying the Langevin equation \cite{Lietal2010Brownian,Gieseleretal2012},
\begin{equation}
\ddot{x}(t)+\Gamma_0\dot{x}(t)+\Omega_0^2x(t)=\frac{1}{m}F_{fluct}(t).
\end{equation}
Here $\Gamma_0$ is the mechanical damping caused by collisions with the surrounding gas, $\Omega_0$ is the trap's natural frequency, and $F_{fluct}$ is a random Langevin force satisfying $\langle F_{fluct}(t)F_{fluct}(t')\rangle=2m\Gamma_0k_BT\delta(t-t')$. The corresponding power spectral density (PSD) of this motion is described by,
\begin{equation}
PSD(\omega) \propto \frac{\Gamma_0}{(\Omega_0^2-\omega^2)^2+\omega^2\Gamma_0^2}.\label{eq:PSD}
\end{equation}

For small displacements, the trap frequency is approximated by $\Omega_0=\sqrt{\frac{k_{trap}}{m}}$, where $k_{trap}$ is the trap stiffness. In the transverse direction, \cite{Gieseleretal2012}
\begin{equation}
k_{trap}=4\pi^3\frac{\alpha P}{c\epsilon_0}\frac{(\mathrm{NA})^4}{\lambda^4},
\end{equation}
depends on the optical system's NA, trapping power, $P$, wavelength, $\lambda$; as well as the particle's polarizability, $\alpha$. For small spherical particles, $\alpha\propto R^3$. Thus the natural frequency,
\begin{equation}
\Omega_0 \propto \sqrt{\frac{R^3}{m}} \propto \sqrt{1/\rho},\label{eq:omega}
\end{equation}
depends density, $\rho$, and is independent of particle size. We can therefore fit a measured PSD (cyan curve in Fig. \ref{fig:position}b) using Eq. (\ref{eq:PSD}), while fixing $\Omega_0$ according to Eq. (\ref{eq:omega}). Determining $\Gamma_0$ from the fit allows us to estimate the particle radius, $R$, by using the Stokes friction coefficient, $\gamma=6\pi\eta R$,
\begin{equation}
\Gamma_0=\frac{\gamma}{m}=\frac{6 \pi \eta R}{\rho(\frac{4}{3}\pi R^3)}=\frac{9 \eta}{2 \rho R^2}.
\end{equation}
Here $\eta$ is the dynamic viscosity of the surrounding medium. The data in Fig. \ref{fig:position}b result in an estimated particle radius of 38 nm, which is within the tolerance quoted from the manufacturer.

\begin{figure}
\begin{centering}
\includegraphics[width=8.5cm]{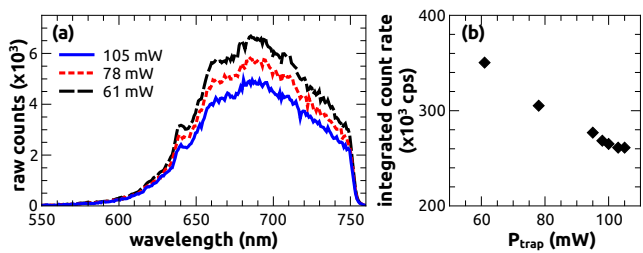}
\caption{(a) Photoluminescence spectra show an increase when the trapping power is reduced, but little change in shape. (b) The spectrally integrated PL rate decreases with increasing trap laser power. Spectra and position measurements (Fig. 2) were measured on the same levitated nanodiamond.  \label{fig:CW}}
\end{centering}
\end{figure}

Photoluminescence spectra from a trapped nanodiamond are shown in Fig. \ref{fig:CW}a. An average excitation power of 37 $\mu$W was used for all measurements, and an exposure time of 5 seconds was used to take each spectrum. Data are plotted for three different trapping powers. The spectra are readily identifiable as the characteristic 637 nm zero phonon line (ZPL) of the NV$^-$ defect state, and broad phonon-assisted sidebands extending beyond 700 nm (truncated at 750 nm due to filtering).  The spectra show an increase in photoluminescence rate across the entire phonon assisted band as trapping power is decreased. Notably, we see no evidence for the NV$^0$ charge state when a continuous trap is used. Fig. \ref{fig:CW}b shows the decrease in spectrally integrated count rate as a function of $P_{trap}$. Quenching of NV photoluminesence has been induced by heating samples to several hundreds of $^{\circ}$C \cite{Plakhotnik2010PLquenching,Toylietal2012SpinLifetime}. The dependence displayed in Fig. \ref{fig:CW}b, however, saturates at higher laser powers, and is therefore inconsistent with heating due to absorption of the trap beam. Rather, it is consistent with a recent study of 2-color excitation of nanodiamonds containing single NV$^-$ centers \cite{Geiselmannetal2013quench1064}.

\begin{figure}
\begin{centering}
\includegraphics[width=8.5cm]{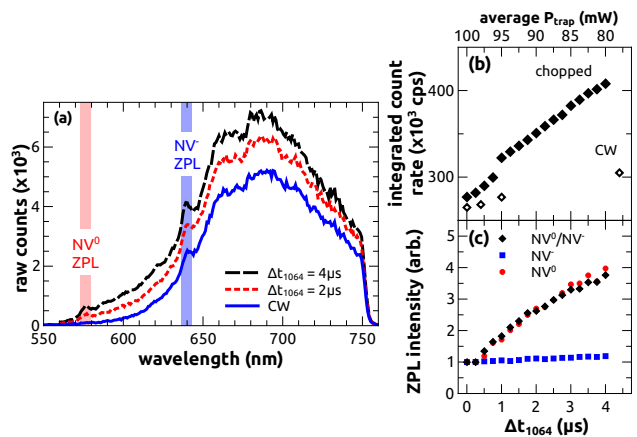}
\caption{(a) Chopping the trap laser at 50 kHz results in the emergence of the neutral (NV$^0$) charge state, as is evidenced by the zero phonon line at 575 nm, and increased counts at wavelengths shorter than 650 nm. As the trap-off duration, $\Delta t_{1064}$, increases, both (b) the total count rate and (c) the relative intensity of the NV$^0$ ZPL increase. The relative intensity of the NV$^-$ ZPL appears to increase very slowly, however this is likely due to a contribution from the rapidly increasing NV$^0$ sideband. Measurements were made on the same diamond as Fig. 2 and 3. \label{fig:Chopped}}
\end{centering}
\end{figure}

Given that reducing the power of the trap laser increases PL yield, we investigated the effect of chopping the trap beam (on the same nanodiamond). A constant trap-on power of 100 mW was used during chopping experiments. We used an AOM to chop the trap at a rate of 50 kHz (cycle period $T=20$ $\mu$s), and varied the trap-off duration, $\Delta t_{1064}$, from 250 ns to 4 $\mu$s. Spectra for $\Delta t_{1064}=0$, 2, and 4 $\mu$s are plotted in Fig. \ref{fig:Chopped}a. The excitation power (37 $\mu$W) was the same as for the continuous trap experiment. We again measure spectra using 5 second exposures, and are thus averaging trap-on and trap-off PL. The spectra in Fig. \ref{fig:Chopped}a show higher count rates at all wavelengths when the trapping laser is chopped. A small peak at 575 nm (NV$^0$ ZPL) and elevated counts below 650 nm clearly indicate the emergence of the NV$^0$ state in the chopped-trap spectra. In Fig. \ref{fig:Chopped}b we plot the spectrally-integrated counts for various values of $\Delta t_{1064}$ (solid diamonds). The PL rates increase with the increase in trap-off time. To compare these with the data in Fig. \ref{fig:CW}b, we compute average trap powers (top horizontal scale) by multiplying the trap-on power (100 mW) by the chopping duty cycle. We plot the integrated count rates from Fig. \ref{fig:CW}b for corresponding average trap powers (open diamonds). Chopping the trap beam results in a much more pronounced increase in integrated count rate than simply reducing the trap power. 

In Fig. \ref{fig:Chopped}c we compare the relative changes in ZPL intensities. Plotted are the the heights of the NV$^-$ ZPL (blue squares) and NV$^0$ ZPL (red dots) relative to the spectral peak ($\sim$690 nm), and the ratio of the intensities of the NV$^0$ ZPL to the NV$^-$ ZPL (black diamonds). In each case, the data are normalized to their CW ($\Delta t_{1064}=0$ ns) values. Clearly, the NV$^0$ ZPL is enhanced as the trap laser is modulated, however, it is not clear from these measurements whether this is due to a suppression of NV$^0$ population, or simply a non-radiative quenching process which preferentially affects NV$^0$.

A number of authors have investigated NV photochromism using single color \cite{Weeetal2007JPCA,Aslametal2013NJP} excitation. Ivanov et al. \cite{Ivanovetal2013OL} have recently reported preferential NV$^-$ excitation by 2-photon absorption in the near infrared (1040 nm), and complete suppression of NV PL attributed to heating has been demonstrated under intense 1064 illumination \cite{Laietal2013arxiv}. To our knowledge, the suppression of NV$^0$ PL under simultaneous excitation by 532 nm and moderate 1064 nm illumination has not been previously reported. A more complete understanding of the mechanism underlying this suppression is important, given the significant implications for studies in which PL from the NV$^0$ state is undesirable.

In conclusion, we have, for the first time, demonstrated nitrogen vacancy photoluminescence from a diamond levitated in a free-space optical trap. Photoluminescence was excited by a 532 nm pulsed laser aligned confocally with the 1064 nm CW trap laser. PL rates are shown to increase as 1064 power decreases, and the nature of this dependence suggests the underlying cause is not thermally driven, indicating that heating is not a significant issue. We observe a high degree of suppression of PL from the NV$^0$ defect state while the trap laser is on. Chopping the trap laser on microsecond timescales increased the overall PL, and resulted in a mixing of PL from both the NV$^0$ and NV$^-$ charge states.

\smallskip
The authors thank B. Deutsch and J. Cosentino for assistance with experiments, and three reviewers for their helpful comments. L.P.N. acknowledges support from the University of Rochester Department of Physics and Astronomy. A.N.V. acknowledges support from the Institute of Optics at the University of Rochester. Capital equipment available for this research was partially funded under Contract Number: W911NF0910425 (N. George, PI) supported by Dr. Richard Hammond, Physics Division, U.S, Army Research Office. J.G. and R.Q. acknowledge support from the European Community's Seventh Framework Program under grant ERC-Plasmolight (no. 259196) and Fundaci\'{o} privada CELLEX. L.N. acknowledges support from the U.S. Department of Energy (Grant No. DE-FG02-01ER15204).

\end{document}